# Rule based Approach for Word Normalization by resolving Transcription Ambiguity in Transliterated Search Queries

Varsha M. Pathak
Research Coordinator
KCES's Institute of Management & Research, Jalgaon
varsha.pathak@imr.ac.in

Manish R. Joshi
Professor, School of Computer Sciences,
Kavayatri Bahinabai Chaudhari North Maharashtra University, Jalgaon India
joshmanish@gmail.com

**Abstract:**

Query term matching with document term matching is the basic function of any best effort Information Retrieval models like Vector Space Model. In our problem of SMS based Information Systems we expect common people to participate in information search. Our system allows mobile users to formulate their queries in their own words, own transliteration style and spelling formation. To achieve this flexibility we have resolved the term level ambiguity due to inherent transcription noise in user query terms. We have developed a rule based approach to select most relevantly close standard term for each noisy term in the user query. We have used four different versions of the rule based algorithm with variation in the rule set. We have formulated this rule set including the basic Levenshtein's minimum edit distance algorithm for term matching.

This paper presents the experiments and corresponding results of Marathi and Hindi language literature information system. We have experimented on Marathi and Hindi literature which include songs, gazals, powadas, bharud and other types in a standard transliteration form like ITRANS.

## 1. Introduction

Mobiles have triggered and expedited the multi-modal communication technology. In modern age it has become an important gadget irrespective of the socio-economic boundaries, [10]. In mobile industry the Short Message Service (SMS) is the most reachable and affordable telecommunication service [2]. It is popularly accepted as the means of person to person information exchange. The statistical data reports as available on [3], citizens of United States sent 69,000 texts every second in 2012" [CTIA 2012]. Similar scenario is observed in most of the developing and developed countries. According to a report of [Connect Mogul], only 43% of smart phone users make calls where as 70% users generally text. It gives convenience to communicate with people even if they are not available to answer immediately. Another interesting point is of response time. According to [CTIA 2010], on an average 90 minutes is the response time for an email, where as a text is answered by the recipient in only average 90 seconds.

Telecom Regulatory Authority of India (TRAI) has recognized Mobile based Value Added Services (MVAS). We referred the guidelines of TRAI on MVAS. TRAI recommends that SMS based Person to Application communication can turn into number of fruitful MVAS. This includes M-Governance, M-Education, M-Banking, M-health as major application areas. ASSOCHAM had forecasted that by the year 2015, in India alone the market of MVAS will reach up to Rs. 482 billion and large part of it can be accommodated by non-voice (text/images) messaging [1].

### 1.1 SMS based Information Retrieval

We contribute to these recent developments by initiating development of a "Mobile based Quick Reference Library", named as "MQuickLib System". It is developed to deal with the related problems to enable users to retrieve literature information by using the basic facility of Mobiles viz. Short Message Service (SMS). The idea is to allow people to send SMS based queries to retrieve required information. The SMS based information retrieval could be considered as the modern implementation of the conventional Information Retrieval methodology. The underline literature shows that SMS based Frequently Asked Question Answer and SMS based Natural Language Flexible Query processing are the research extensions to the existing Information Retrieval. The wide coverage of Mobiles at grass roots of the society demands to explore **Information Retrieval** models with a new personalization dimension. Thus MQuickLib is designed to allow users to form their queries in their own natural language by using their own style of spelling the terms. To overcome the scripting problems of languages across the globe, transliterated text is allowed to the users. The users can spell the words of their language in Roman script. We have trained and tested the system on Marathi and Hindi languages. The related problems are investigated with a systematic development of an experimental model. The experiments conducted have revealed number of problems. The problems like dealing with transcription ambiguity and Named Entity Recognition (NER) ambiguities are the major issues those are investigated in the research undertaken by us. From the corresponding experiments a systematic model of Relevance Feedback Mechanism has surfaced out. Relevance Feedback Mechanism is the most natural way of adapting with the user's information requirement, their style of formulating the queries by using the acronyms, synonyms, abbreviations and SMS shortcuts.

We have explored different important IR techniques in **MQuickLib** with proven good performance result. The Vector Space Model (VSM), is the basic IR model. This model is developed with enhanced features to support the RFM strategy. The Relevance Feedback Mechanism developed follows a hybrid approach to resolve the ambiguities at various stages of system's responses for user queries.

## 1.2 Transcription ambiguity

Transcription ambiguity occurring in user's query due to flexible styles of transliteration is the upcoming problem due to popularization of SMS in native languages. Many users use their own style for formulating the SMS query in their native language for Information Retrieval.

In our problem in order to use transliterated queries for Marathi Literature access on Mobiles we need to handle transcription ambiguity. The end users apply non-standard transliteration while compiling their information search queries in their own language. They use mixed language to formulate their queries.

This paper presents the work on SMS based Literature Information System as the extension to digitalization of present library facilities. This service needs to resolve the noisy terms to possible normal terms those occur in the systems vocabulary. We identify this as a word normalization problem. The systems vocabulary thus need to be constituted of the normal terms occurring in the literature. In our problem we have constructed the vocabulary by processing the ITRANSed Literature of Marathi language available on the internet sites like https://www.sanskritdocuments.org and http://www.giitaayan.com etc. This vocabulary of Marathi literature words extracted from this available contents is built by using the customized the Vector Space Model for this domain [11].

## 1.3 ITRANSed Documents

The system accepts queries in users own vocabulary with flexible transliteration style. Marathi/Hindi language use Devanagari as its natural script, whereas people use English alphabets to spell the Marathi/Hindi words while interacting with others on computer network on mobile network. Thus we assume that users enter roman scripted strings to formulate Marathi/Hindi queries to access literature information. The counter part of the problem is that the Literature documents are in standard transliterated form like ITRANS. The Vector Space constructed by processing these documents forms a standard vocabulary of literature terms of respective language. To allow flexibility in Romanization of literature queries the system needs to enhance its ability to identify the query terms correctly by matching them with the most relevant standard vocabulary term which we term as "ITRANSed Normal terms". This problem is considered as the spelling correction problem. If a term t is spelled incorrectly it may not occur in system's vocabulary V, such term is defined as noisy term. We need to map this noisy term to a standard term which occurs in the system's vocabulary.

## 2. Problem Modeling

We compared this problem with the SMS based Frequently Asked Question Answering system attained by other researchers. Number of researchers came with new ideas to solve the SMS normalization problem in effective manner. Forum for Information Retrieval and Evaluation (FIRE) has raised number of related research topics in their tracks assigning challenges to resolve the query translation, Query normalization in European and also in Indian languages like Marathi, Hindi, Bengali and many others. The researchers [8], [9] have worked on this problem by applying Levenshtein's algorithm to compute distance between two strings in alphabetical manner. This Levenshtein's algorithm follows the dynamic programming method like Longest Common Subsequence.

The important contribution of our model is that instead of normalizing the user query by using predefined set of Normal Queries in the system, our model normalizes individual terms by selecting a set of Normal Terms from the system's vocabulary. These terms are at closer edit distance from the query term. For each Noisy term $t \in Q$ and a Normal Term $v \in V$ the minimum edit distance $ed = t \simeq v$ is computed by applying the Levenshtein's algorithm [8].

## 2.1 Minimum Edit Distance

Edit distance between two strings is computed to measure the number of edit operations required to make the two strings exactly equal. The edit operations include insert, delete and substitute character in a string. Any two valid strings can have variable edit distance. That means a string can be edited to another by using different way of editing. For example the strings "RISK" and "MASK" has edit distance 2, as two substitute operations viz R/M and I/A can modify "RISK" to "MASK". Similarly edit distance between Marathi words as in eg. 2 "phule" and "phulen" is 1, as single insert (of character n) operation in "phule" converts it to string "phulen". Another strings in eg. 3, "phulen" and "fulaat" are at edit distance 5 as well as 4. Thus minimum edit distance between such terms is calculated in this case. The minimum edit distance follows dynamic programming method. It applies to find minimum number of edit operations required to modify a string S1 to another string S2. In above example the minimum edit distance between "phulen" and "fulaat" is four. The minimum edit distance is applicable not only to find distance between two valid strings in certain vocabulary, but also could be used to find the closest valid string occurring in a vocabulary for a corrupted string. Mostly it is applied in spell correction algorithms. We used this method with a set of rules identified specially for query term normalization.

## 2.2 Query Term Normalization

In query term normalization problem we applied the basic Levenshtien's minimum edit distance function to map a noisy term in a user query to a normal term that occurs in the system's vocabulary. A few examples of the minimum edit distance between words for a few query terms are illustrated in Figure-1. The system's vocabulary is nothing but the inverse document term vector that we have generated from the Marathi and Hindi literature's document corpus to formulate these rules. A set of more than hundred and fifty Marathi queries and forty Hindi song queries are used to train the system by varying these rules. Finally these tested rules are combined to formulate our "Rule based Query Term Normalization Model" for the MQuickLib system. This model formulates a Proxy Weight PW for a normal term with respect to a noisy term, by applying following mathematical model.

The basic edit distance formulation is applied along with a set of few rules. These rules are defined by considering the possible noise induced by users while transliterating a Marathi or Hindi query on the mobile interface. The practices applied by users while typing a query in SMS form is taken into account

For a term t ⊂ Q, a normal term set NT is formed by applying a set of rules that are explained below. The normal term set NT is thus a subset of vocabulary V.

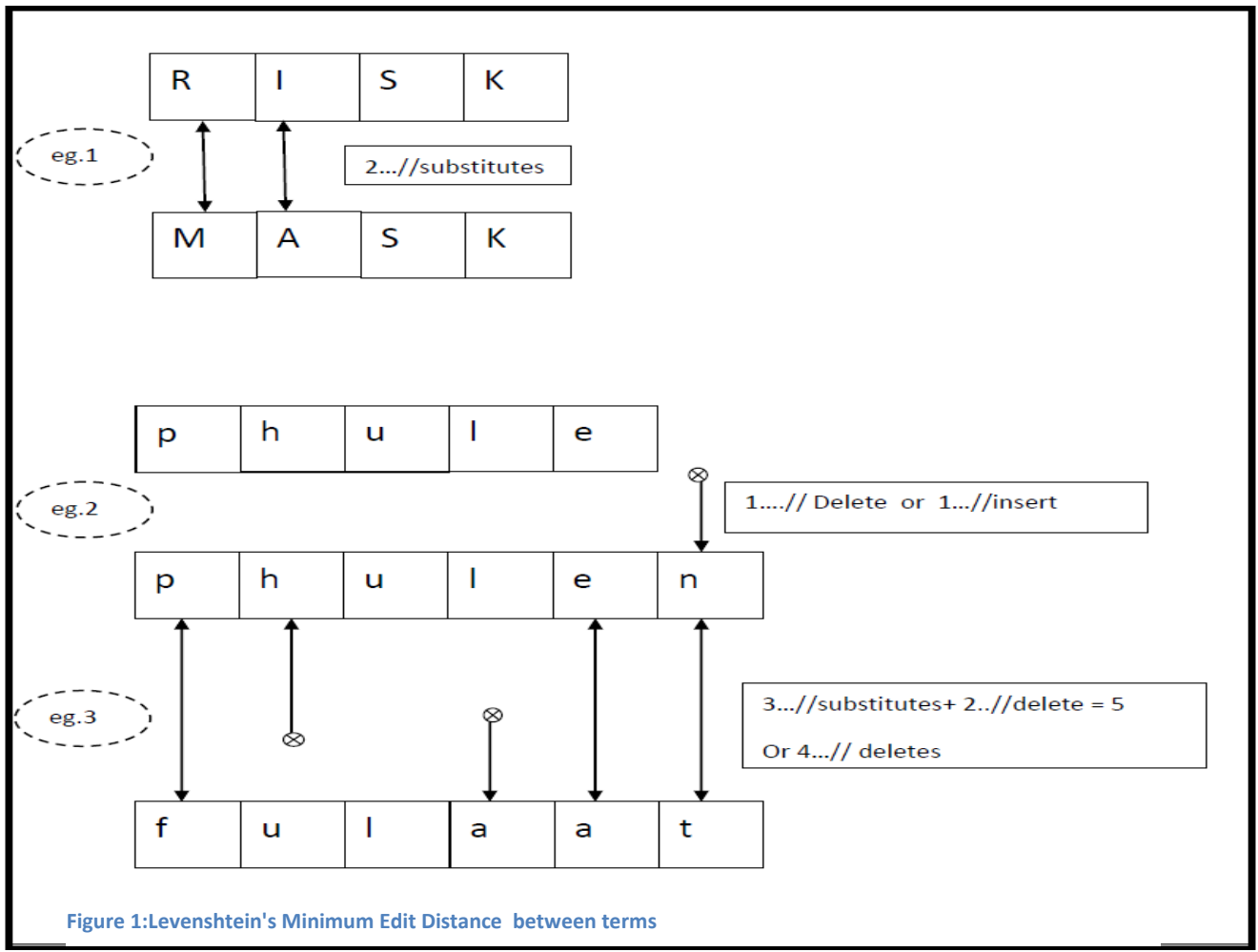

Figure 1: Levenshtein's Minimum Edit Distance between terms

i.e. ti ∈ Q a normal term ntj ∈ NT ⊂ V is candidate normal term used to search the relevant document for the user query Q. For the candidature of a term vj in the set NT following a set of rules are experimented with basic Minimum Edit Distance function. For each pair of noisy term and normal term, the basic formula applied is as given in Eq. 1. As minimum edit distance between two terms measures how closer the two terms are, the ratio of the edit distance to the length of longest term quantifies the comparison of closeness of two or more terms from a specific term. If a term t is at edit distances d1 and d2 from two terms x and y with length l1 and l2 respectively then if d1/l1 < d2/l2 then x is at closer distance from t as compared to y from t. With this mathematical rule the proxy weight of a term v ∈ V quantifies its similarity with a user query term t with Eq. 1. This formulation is modified to smoothen the similarity measure in case if the distance is 0. This variation is given as in Eq. 2.

$$ProxWt_{t,v} = PrunWt - \frac{ed_{t,v}}{len} \qquad Eq.1$$

$$ProxWt_{t,v} = PrunWt - \frac{ed_{t,v} - 1}{len} \qquad Eq.2$$

Where..

_ PrunWt is the term weight calculated by applying a weighting rules $R_w \subset R$

_ ed = minDist(t,v) calculated by Levenshtien's minimum edit distance algorithm[9].

_ len is computed applying term length related rules $R_l \subset R$.

The set of rules R is defined and discussed as follows.

## 3. Rules and Models

Following preconditions are considered for the framework of the models varied by set of rules.

• The SMS query Q is the sequence of n number of terms $t_1$, $t_2$, …. $t_n$. It is clear that the query is noisy as one or more terms are distorted due to noise in transcription and SMS encoding mechanisms.

• V is the system's vocabulary computed by the Vector Space Mechanism on the set of literature documents. Let the terms $v_1$, $v_2$…… $v_m$ are the normal terms as they occur in this set V.

• ed = $t_i \simeq v_j$ = minDist($t_i$, $v_j$) is the function that calculates minimum edit distance between two terms $t_i$ and $v_j$.

• Let $W_1 = 0:60$, $W_2 = 0:40$, $W_3 = 0:2$, $W_4 = 0:75$ and $W_5 = 0:25$ are the weights used in the rules. These weights are computed for fine tuning the term normalization. The weights are stabilized by varying them from 0 to 1 through number of experiments with the objective of associating relevant normal terms with the noisy terms.

Following are the set of term length rules defined in the experiment. For these set of rules, let l1 and l2 are the lengths of two string $v_j$ and $t_i$ respectively.

Threshold condition is ed < α To compute Proxy Weight for the term $v_j$, with respect to query term ti, the threshold value of edit distance ed($t_i$, $v_j$) is equal to half of the len, where len is the term length calculated applying rule set Rl ⊂ R.

It means if edit distance goes beyond this threshold function value then the respective vocabulary term $v_j$ is not included in Candidate Normal Term set NT.

The rules set for proxy weight calculation are designed on the basis of the length of tokens to match and a few rules are defined for matchmaking of the tokens.

## 3.1 Comparison of Models

By selecting different combinations of above rule set we have identified different versions of our model. The best performing four models on the basis of their Pruning Weight scores are selected for finalizing the matchmaking function of Proxy Similarity. We define them in the form of different models. By comparing their performance on the basis of Precision Recall we have selected four variations of these models. These models and their respective rules are described below.

**Table 1: Rules Set**

| Length Related Rules | |
|---|---|
| Rule 1 | ($l_1 > 1$ and $l_2 > 1$) shall be true is essential precondition for proxy weight calculation |
| Rule 2 | Compute len = $l_1 > l_2$ ? $l_1$ : $l_2$ |
| Rule 3 | For a vocabulary term $v_j \subset V$, if $l_1 > l_2$ then len = $l_1$, and $l_1 > l_2$ is necessary condition to qualify vocabulary term (Standard term) $v_j$ for further process of Proxy Weight calculation. |
| Rule 4 | Compute len = ($l_1 + l_2$) /2 |
| Rule 5 | Let length threshold α = *len/2* i.e. half the len value the minimum edit distance then only consider the term for Proxy Weight calculation. |
| **Character Matching Rules** | |
| Rule 6a | if first characters of $t_i$ and $v_j$ match then $flag_1 = 1$, $wf_1 = wt_1$ |
| Rule 6b | if first characters of $t_i$ and $v_j$ do not match then $wf_1 = wt_2$ |
| Rule 7a | if second characters of $t_i$ and $v_j$ match then $wf_2 = wt_2$ |
| Rule 7b | if second characters of $t_i$ and $v_j$ do not match then $wf_2 = wt_3$. |
| Rule 8a | if last consonant characters of the terms $t_i$ and $v_j$ match then $wf_3 = wt_4$. |
| Rule 8b | if last consonant characters of the terms $t_i$ and $v_j$ do not match then $wf_3 = wt_5$. |
| Pruning Weight = $wf_1 + wf_2 + wf_3 + wf_4 + wf_5$. | |

1. Model 1: Basic Levenshtein's Model – This basic model assigns weights to the standard terms computed by applying Eq. 3. Here edv is computed by applying Levenshtein's minimum edit distance between the noisy term t and standard term v. No other additional constraint is used in this model. Thus the pruning weight is set to 1. Similarly len is the length, computed as per the length specific Rule 2 and 5.

$$ProxWt_{t,v} = 1 - \frac{ed_{t,v}}{len} \qquad Eq.3$$

2. Model 2 : First Letter Match Model – In this model the terms are ranked on the weights by adding the constraints based on the weight rule set including rules 6a, 6b, 7a, 7b which checks if first two letters match or not and assigns weights accordingly. Length factor len is computed by applying Rule 4 and 5.

3. Model 3: Last Consonant Match- In this model in addition to first two characters the last consonant is matched. The Pruning weights are computed accordingly by applying rules 6a, 6b, 7a, 7b, 8a and 8b those are Length specific rules same as in Model 2.

4. Model 4: Short Term Match Model - The Model-3 has proved better working on the sample data. It is further improved by restricting the length specific rules by applying the rules Rule 3 and Rule 5.

## 4. Results Analysis

The above four versions are evaluated by applying standard measure to select the final model in actual implementation of the system. The experiments are performed on the word samples those show significant noise in the user queries. The Hindi data set of 108 queries has produced of 56 noisy word and from Marathi data set of 152 queries has produced near about 68 noisy words. These noisy word samples are used in the evaluation of the above four normalization models. These samples show significant variation in mapping a noisy term to standard terms. The results are analyzed by applying standard measures as discussed below. The best performing model is then used for the development of the MQuickLib system.

### 4.1 Evaluation Measures-

For comparison of these models four standard measures are used. The Mean Reciprocal Rank (MRR), Precision for top 1 resultant term (p@1), Precision for top five resultant terms (p@5) and Precision for top 10 resultant terms (p@10) are computed for all sample terms. Average of each of these values are computed on all sample terms. The Table-2 shows the results of the Average values of these standard measures on the sample data sets. The respective average values of these measures Avg_MRR_h, Avg_PH_1, Avg_PH_5 and Avg_PH_10 are for Hindi data set. Similarly Avg_MRR_m, Avg_PM_1, Avg_PM_5 and Avg_PM_10 are the average values on the Marathi data set.

The graphical representation of these results could be analyzed for the comparison of the performance of the models on the sample data sets. We can understand that the Models M3 and M4 are proved to be better performing than M1 and M2.

## 5. Conclusion

Query term matching with document term matching is the basic function of any best effort Information Retrieval models like Vector Space Model. In our problem of SMS based Information System we expect common people to participate in information search. The MQuickLib system allows mobile users to formulate their queries in their own words, own transliteration style and spelling formation. To achieve this flexibility the term level ambiguity due to inherent transcription noise in user query terms is resolved in this system. A rule based approach to select most relevantly close standard term for each noisy term has been developed. This paper presents the result of four different modifications in such rule based model. These models vary in the rule set in the context of information retrieval including the basic minimum edit distance algorithm.

Table 2: Evaluation of Models M1 to M4

| Measure | M1 | M2 | M3 | M4 |
| --- | --- | --- | --- | --- |
| Avg_MRR_h | 0.028333 | 0.29 | 0.916667 | 0.916667 |
| Avg_PH_1 | 0 | 0.166667 | 0.833333 | 0.833333 |
| Avg_PH_5 | 0 | 0.458333 | 0.75 | 0.916667 |
| Avg_PH_10 | 0.091667 | 0.708333 | 1 | 1 |
| Avg_MRR_m | 0.8125 | 0.545 | 0.875 | 0.79125 |
| Avg_PM_1 | 0.75 | 0.375 | 0.875 | 0.75 |
| Avg_PM_5 | 0.79 | 0.77 | 0.79 | 0.76 |
| Avg_PM_10 | 0.80125 | 0.8275 | 0.84875 | 0.86 |

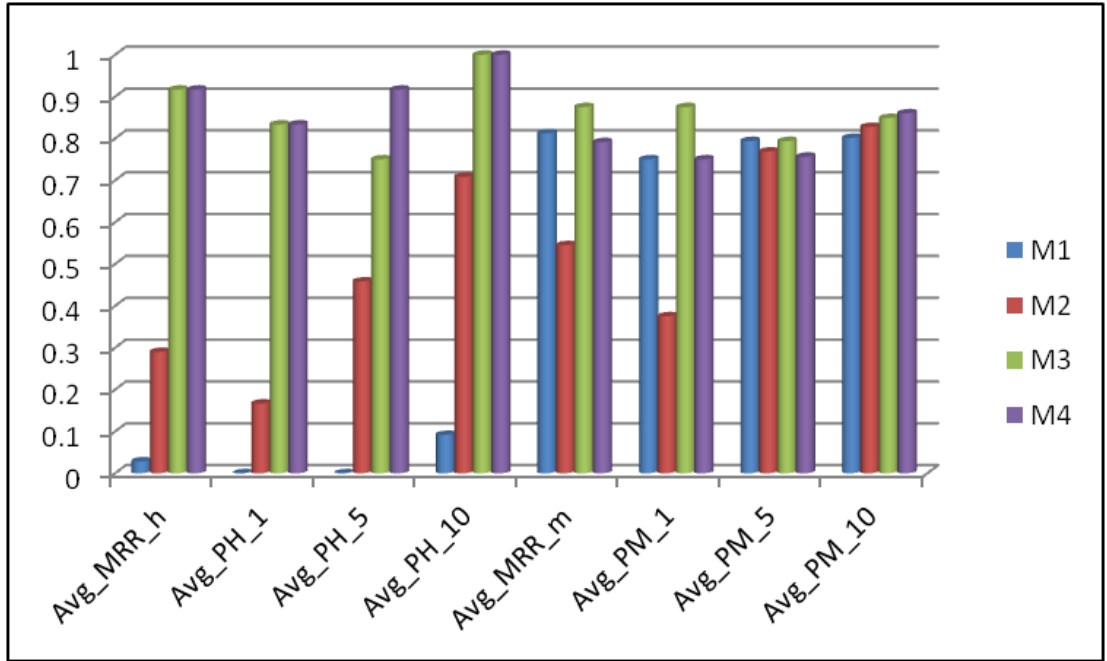

Figure 2: Chart showing Average of Standard Measures on sample Data Sets